\documentclass[aps,prb,reprint,twocolumn,showpacs,floatfix,superscriptaddress,nofootinbib]{revtex4}
\usepackage{graphicx}
\usepackage{amssymb}
\usepackage{amsmath}
\usepackage{lmodern}
\usepackage{color}
\usepackage{hyperref}
\usepackage{empheq}
\usepackage[makeroom]{cancel}
\usepackage{epstopdf}
\begin{document}

\title{Poor man's scaling: anisotropic Kondo and  Coqblin--Schrieffer models}

\author{Eugene Kogan}
\email{Eugene.Kogan@biu.ac.il}
\affiliation{Jack and Pearl Resnick Institute, Department of Physics, Bar-Ilan University, Ramat-Gan 52900, Israel}
\affiliation{Max-Planck-Institut fur Physik komplexer Systeme,  Dresden 01187, Germany}
\affiliation{Donostia International Physics Center (DIPC), Paseo de Manuel Lardizabal 4, E-20018 San Sebastian/Donostia, Spain}

\begin{abstract}
We discuss  Kondo effect  for a general model, describing   a  quantum impurity with degenerate energy levels, interacting with a gas of itinerant electrons, and derive scaling equation to the second order for such a model.
We  show how the scaling equation for the spin-anisotropic Kondo model with the power law density of states  (DOS) for itinerant electrons follows from the general scaling equation.
We introduce  the anisotropic  Coqblin--Schrieffer  model,
apply the general method to derive scaling equation for that model  for the power law DOS, and
integrate the derived  equation analytically.
\end{abstract}


\maketitle

\section{Introduction}

Observed under appropriate conditions  logarithmic increase (with the decreasing of temperature) of the scattering of the itinerant electrons by an isolated magnetic impurity was explained in 1964 in a seminal paper by Kondo, entitled
"Resistance Minimum in Dilute Magnetic Alloy" \cite{kondo}.
Soon after it became clear that the phenomenon  is manifested not only in resistivity, but in nearly all thermodynamic and kinetic properties \cite{hewson}.
The theoretical analysis of this effect turned out to be very fruitful, and led to the appearance of many approaches and techniques, which became
 paradigms in many totally different fields of physics.
One of such approaches was the so called poor man's scaling, pioneered by Anderson \cite{anderson}.

Though initially only magnetic impurity scattering was considered, later it became understood that similar effect can appear in case of
a general quantum impurity. Such a general model was thoroughly reviewed in 1998 by Cox $\&$ Zawadovski \cite{cox}.

Recently the models where the density of states (DOS) of itinerant electrons in the vicinity  of  the Fermi level is the power function of energy
has attracted a lot of interest
\cite{sengupta,wehling,vojta,uchoa,fritz,kogan,ingersent}. The  scaling was generalized to be applicable to such systems in the work by
Withoff and  Fradkin \cite{fradkin}.

Following the long line of works where spin-anisotropic Kondo model was studied \cite{anderson,shiba,yosida,kogan}, we decided to revisit the problem of  scaling equation for the  Kondo effect in general \cite{cox}, and introduce and study the  anisotropic  Coqblin--Schrieffer (CS) model  \cite{coqblin}
in particular. Notice that it is known that the spin anisotropy can substantially change the physics of the Kondo effect in comparison with the isotropic case \cite{cox,costi,irkhin2,thomas}.
 The CS model, though being well studied previously \cite{hewson,rajan,schlottmann,andrei,zlatic,bazhanov,kikoin},  draw a lot of attention recently in connection with the   studies of quantum dots \cite{kuzmenko},  heavy fermions \cite{desgranges} and ultra-cold gases \cite{figueira,avishai}.

The rest of the paper is constructed as follows.
We formulate in Section \ref{second} a  poor man's scaling equation to second order for a general  model,
describing   a quantum impurity embedded into a gas of itinerant electrons.
We also generalize the  equation to the case of
power law behavior of the DOS in the vicinity of the Fermi level.
In Section \ref{spin} we show how the obtained earlier scaling equation for the spin-anisotropic Kondo model
follow from  the general scaling equation.
In Section \ref{gcs} the  $XXZ$ CS model is introduced and  scaling equation for this model is derived.
Then everything is generalized to the case of the  anisotropic CS model.
The scaling equation both in the particular case of  the $XXZ$  CS model and in the general case of the anisotropic CS model are integrated   analytically.
We conclude in Section \ref{conclusions}.  Some mathematical spin-offs are presented in the Appendix.

\section{Perturbation theory}
\label{second}

\subsection{Kondo effect as explained by Kondo}

The  Hamiltonian we start from is \cite{cox}
\begin{eqnarray}
\label{hamilto}
H&=&H_0+V  \nonumber\\
&=&\sum_{{\bf k}\alpha}\epsilon_{\bf k}c_{{\bf k}\alpha}^{\dagger}c_{{\bf k}\alpha}
+\sum_{\substack{{\bf k}{\bf k}'\\\alpha\beta,ab}}V_{\beta\alpha,ba}X_{ba}c_{{\bf k}'\beta}^{\dagger}c_{{\bf k}\alpha},
\end{eqnarray}
where $c^{\dagger}_{{\bf k}\alpha}$ and $c_{{\bf k}\alpha}$ are  electron creation and annihilation operators of itinerant electron with  wave vector ${\bf k}$ and internal quantum number $\alpha$,   $\epsilon_{\bf k}$ is the energy of the electron; $X_{ba}=|b><a|$, where $|a>,|b>$ are the internal states of the scattering system, is the Hubbard $X$-operator.

In this paper we use the old-fashioned on-the-energy-shell perturbation theory \cite{landau}  following the paper by Kondo; similar approach was applied to the Anderson model by Haldane \cite{haldane}. For the Hamiltonian (\ref{hamilto})
the transition probability per unit time from the initial state of the whole system $m$ to the final
state $n$ is given to the second Born approximation by  \cite{kondo}
\begin{eqnarray}
\label{pertn}
W_{n\leftarrow m}=2\pi\delta(E_n-E_m)|T_{nm}|^2,
\end{eqnarray}
where
\begin{eqnarray}
\label{per}
T_{nm}=V_{nm}+\sum_{\ell}\frac{V_{n\ell}V_{\ell m}}{E_m-E_{\ell}}+\dots
\end{eqnarray}
is the scattering matrix.
Explicitly Eq. (\ref{per})  takes the form \cite{kondo}
\begin{eqnarray}
\label{perturbation2}
T_{\beta\alpha,ba}=V_{\beta\alpha,ba}
+\sum_{\gamma,c}V_{\beta\gamma,bc}V_{\gamma\alpha,ca}\sum_{\epsilon_{\bf q}>0}\frac{1}{\epsilon-\epsilon_{\bf q}}\nonumber\\
-\sum_{\gamma,c}V_{\gamma\alpha,bc}V_{\beta\gamma,ca}\sum_{\epsilon_{\bf q}<0}\frac{1}{\epsilon_{\bf q}-\epsilon}.
\end{eqnarray}
Writing Eq. (\ref{perturbation2}) we assumed that
in the state $m$  all the electron states  below the Fermi surface (corresponding to $\epsilon_{\bf k}=0$) are occupied, and there is an additional electron with the
wave vector ${\bf k}$ and internal quantum number $\alpha$; the scattering system is in the state $|a>$. In the state $n$  all the electron states  below the Fermi surface are again occupied, and there is an additional electron with the
wave vector ${\bf k}'$ ($\epsilon_{{\bf k}'}=\epsilon_{\bf k}=\epsilon$) and internal quantum number $\beta$; the scattering system is in the state $|b>$.

In the R.H.S. of Eq. (\ref{perturbation2}) the second term describes the processes when the electron with ${\bf k}\alpha$ is first scattered to the unoccupied state ${\bf q}\gamma$ and
then to ${\bf k}'\beta$, and the third term describes the processes when  an
electron from an
occupied state ${\bf q}\gamma$ is first scattered to ${\bf k}'\beta$ and
then the electron with ${\bf k}\alpha$ fills up the state ${\bf q}\gamma$ which is now empty \cite{kondo}.

Equation (\ref{perturbation2}) clearly explains the connection between  the dynamics of the scattering system and the Kondo effect. For static impurity, Eq. (\ref{per}) takes the form
\begin{eqnarray}
\label{perturbation3}
T_{\beta\alpha}=V_{\beta\alpha}
+\sum_{\gamma,{\bf q}}\frac{V_{\beta\gamma}V_{\gamma\alpha}}{\epsilon-\epsilon_{\bf q}}+\dots.
\end{eqnarray}
Because all the  integrals with respect to energy in perturbation series terms are understood in the Principal Value sense,  the denominator   going to zero
is by itself not a  problem, and
the second order terms just gives a correction to the matrix element of the order of the ratio of the scattering energy $V$ to the band width  (we consider electron band $\epsilon\in [-D_0,D_0]$ and assume that the ratio is $V/D$ is small). In addition, this correction is only weakly $\epsilon$-dependent.
On the other hand, each of the second order terms in Eq. (\ref{perturbation2}) contains large logarithmic multiplier $\ln(\epsilon/D_0)$,
because of  strongly asymmetric range of integration \cite{hewson}; due to the existence of the impurity quantum numbers, these terms do not add up to Eq. (\ref{perturbation3}).

\subsection{Scaling equation to second order}
\label{s2}

Equation (\ref{perturbation2}), as it is written down, allows to calculate  scattering at high temperatures. However, it allows
more -- to obtain a scaling equation for the Kondo model in the framework of the approach
 pioneered by Anderson \cite{anderson,hewson}, which allows  to selectively sum up the infinite perturbation series, using explicitly only the first two terms of such an expansion, as presented in Eq. (\ref{per}).

We are interested only in the matrix elements  between the electron states at a distance from  the Fermi energy much less than the band width.
The brilliant idea of Anderson, applied to the present situation, consists in reducing the band width  of the itinerant electrons
from $[-D_0,D_0]$ to $[-D_0-dD,D_0+dD]$ ($dD<0$)
 and taking into account the terms which corresponded to summation  in Eq. (\ref{perturbation2})
 with respect to the electron states in energy intervals $[-D_0,-D_0-dD]$ and $[D_0+dD,D_0]$ by renormalizing $V_{\beta\alpha,ba}$.
Notice that such renormalization can be performed provided that $|\epsilon|\ll D_0$ and hence can be discarded in the denominators of the second order terms in Eq. (\ref{perturbation2}).
In our particular case, like  in general, renormalization is the reduction of  the Hilbert space ${\cal H}$ and changing the Hamiltonian $H$ so as  to keep the physical observable $T$ constant.
Thus we obtain
\begin{eqnarray}
\label{scalin}
dV_{\beta\alpha,ba}=\rho\sum_{\gamma,c}\left[V_{\beta\gamma,bc}V_{\gamma\alpha,ca}
-V_{\gamma\alpha,bc}V_{\beta\gamma,ca}\right]\frac{dD}{D_0},
\end{eqnarray}
where  $\rho$ is the density of states of itinerant electrons (assumed to be constant).

Poor man's scaling consists in changing Eq. (\ref{scalin}) to
\begin{eqnarray}
\label{scaling0}
&&dV_{\beta\alpha,ba}(D)\\
&&=\rho\sum_{\gamma,c}\left[V_{\beta\gamma,bc}(D)V_{\gamma\alpha,ca}(D)
-V_{\gamma\alpha,bc}(D)V_{\beta\gamma,ca}(D)\right]\frac{dD}{D},\nonumber
\end{eqnarray}
where $D$ is now a running parameter.
From Eq. (\ref{scaling0}) we obtain the scaling equation
\begin{eqnarray}
\label{scaling}
\frac{dV_{\beta\alpha,ba}}{d\ln\Lambda}=\rho\sum_{\gamma,c}\left[V_{\beta\gamma,bc}V_{\gamma\alpha,ca}
-V_{\gamma\alpha,bc}V_{\beta\gamma,ca}\right],
\end{eqnarray}
where $\Lambda=D/D_0$ (actually, the change of (implied) argument of each matrix element $D\to\Lambda$ in Eq. (\ref{scaling}) was done for no reason),
and, of course, Eq. (\ref{hamilto}) should be now understood as
\begin{eqnarray}
\label{hamilto2}
H=\sum_{{\bf k}\alpha}\epsilon_{\bf k}c_{{\bf k}\alpha}^{\dagger}c_{{\bf k}\alpha}
+\sum_{\substack{{\bf k}{\bf k}'\\\alpha\beta,ab}}V_{\beta\alpha,ba}(\Lambda)X_{ba}c_{{\bf k}'\beta}^{\dagger}c_{{\bf k}\alpha}.
\end{eqnarray}

Let the matrix $V_{\beta\alpha,ba}$ is presented as
a sum of direct products of matrices, acting in $ab$ and $\alpha\beta$ spaces respectively
\begin{eqnarray}
\label{product}
V=\sum_{p\pi}G_p\otimes \Gamma_{\pi}c_{p\pi},
\end{eqnarray}
where the set of matrices $\{G_p\}$ is closed with respect to commutation and hence generates some Lie algebra $g$; so is the set of matrices $\{\Gamma_p\}$ (Lie algebra $\gamma$).

With the help of Eq. (\ref{product}) we can write down Eq. (\ref{scaling})  in a more transparent form
\begin{eqnarray}
\label{scaling25}
\sum_{p\pi}G_p&\otimes &\Gamma_{\pi}\frac{dc_{p\pi}}{d\ln\Lambda}\nonumber\\
=\frac{1}{2}\rho\sum_{s\sigma t\tau}\left[G_s,G_t\right]&\otimes& \left[\Gamma_{\sigma},\Gamma_{\tau}\right]c_{s\sigma}c_{t\tau}.
\end{eqnarray}
Introducing structure constants of the
 Lie algebras $g$ and $\gamma$  as $f^p_{st}$ and $\varphi^{\pi}_{\sigma\tau}$
\begin{eqnarray}
\left[G_s,G_t\right]=i\sum_pf^p_{st}G_p,\;\;\;
\left[\Gamma_{\sigma},\Gamma_{\tau}\right]=i\sum_{\pi}\varphi^{\pi}_{\sigma\tau}\Gamma_{\pi},
\end{eqnarray}
we can write down Eq. (\ref{scaling})  in an even more transparent form
\begin{eqnarray}
\label{scaling2}
\frac{dc_{p\pi}}{d\ln\Lambda}=-\frac{1}{2}\rho\sum_{st\sigma\tau}f^p_{st}\varphi^{\pi}_{\sigma\tau}c_{s\sigma}c_{t\tau}.
\end{eqnarray}

Typically, the  $\{G_p\}$ ($\{\Gamma_{\pi}\}$) appear as infinitesimal operators of some Lie group $G$ ($\Gamma$), and hence are Hermitian.
Actually, this can be said the other way round. Assuming that the matrices  $\{G_p\}$ ($\{\Gamma_{\pi}\}$) are Hermitian, we see that the algebra $g$ ($\gamma$) is real, and, hence, by Lie's third theorem is the Lie algebra of some simply connected Lie group \cite{hall}.
Anyway, if  $\{G_p\}$ ($\{\Gamma_{\pi}\}$) are Hermitian,  the matrix $c_{p\pi}$ is real.

Consider an important particular case  when  $\gamma\equiv g$. (Matrices  $\{G_p\}$ and $\{\Gamma_{\pi}\}$ not necessarily realize the same representation of the algebra, but have the same commutation relations; to emphasize that we will designate $\{\Gamma_{\pi}\}$ as $\{\Gamma_p\}$.)
 If we assume that the matrix $c_{p\pi}$, in addition to being real, is symmetric, it can be diagonalized by a unitary transformation of the generators (such transformation does not change the commutation relations), and the matrix keeps its diagonal form in the process of renormalization.  Thus
Eq. (\ref{product}) can be "reduced to the principal axes", that is to the form
\begin{eqnarray}
\label{pro}
V=\sum_{p}G_p\otimes \Gamma_pc_p,
\end{eqnarray}
and Eq. (\ref{scaling2}) takes the form
\begin{eqnarray}
\label{scaling26}
\frac{dc_{p}}{d\ln\Lambda}=-\frac{1}{2}\rho\sum_{st}(f^p_{st})^2c_{s}c_{t}.
\end{eqnarray}
Equation (\ref{scaling26}) will be solved in Section \ref{spin}. General analysis of the equation for possible three-dimensional Lie algebras will be presented in Appendix \ref{three}.

\subsection{Power law DOS}
\label{power}

The results of the previous Section can be easily generalized  to the case when the electron dispersion law determines the power law dependence of the DOS upon the energy
\begin{eqnarray}
\label{e}
\rho(\epsilon)=C|\epsilon|^{r},\;\;\;\text{if}\;\;|\epsilon|<D_0,
\end{eqnarray}
where $r$ can be either positive or negative \cite{fradkin}. (We  consider in this paper only  particle-hole symmetric  DOS; the influence of high particle-hole asymmetry on Kondo effect was studied  in Ref. \onlinecite{horvat}.) Notice that $r=0$ corresponds to the previously considered case of flat DOS.
In this case  instead of Eq.  (\ref{scaling}) we obtain
\begin{eqnarray}
\label{scaling8}
\frac{dV_{\beta\alpha,ba}}{d\ln\Lambda}=rV_{\beta\alpha,ba}
+G\sum_{\gamma,c}\left[V_{\beta\gamma,bc}V_{\gamma\alpha,ca}
-V_{\gamma\alpha,bc}V_{\beta\gamma,ca}\right],\nonumber\\
\end{eqnarray}
where  $G=CD_0^r$ is the DOS at the original band edges.
Eq. (\ref{scaling8}) is invariant with respect to simultaneous change of sign of $r$,  of all matrix elements of $V$ and of the direction of flow. Hence further on we consider explicitly  only the case of positive $r$.

The appearance of the linear term in the R.H.S. of Eq. (\ref{scaling8}) demands explanation \cite{fradkin,W}.
To get scaling equation, in addition to integrating out the states at the band edges, another procedure is needed, both for the case of flat DOS, and for the case of power law DOS.
We did not mention it in the previous Section, because it does not change the equation, but in the case of power law DOS it does.  After integrating out the states at the band edges we have
to restore the original band width,  which demands  decrease of the unit of energy by a factor of $D_0/D$. To keep the DOS constant we should increase the unit of volume by the  factor
 $(D_0/D)^{r+1}$.  One should understand that in Eq. (\ref{hamilto}), $V$ has units of energy multiplied by volume, so after all the rescalings the perturbation is multiplied by the factor
of $(D/D_0)^r$, which explains the appearance of the linear term.

To make Eq. (\ref{scaling8}) look similar to Eq. (\ref{scaling}) we introduce new variables  $\lambda=\left(D/D_0\right)^r$ and
$\widetilde{V}=V/r\lambda$, after which we can write down Eq. (\ref{scaling8}) as
\begin{eqnarray}
\label{scaling9}
\frac{d\widetilde{V}_{\beta\alpha,ba}}{d\lambda}=G\sum_{\gamma,c}\left[\widetilde{V}_{\beta\gamma,bc}\widetilde{V}_{\gamma\alpha,ca}
-\widetilde{V}_{\gamma\alpha,bc}\widetilde{V}_{\beta\gamma,ca}\right].
\end{eqnarray}

\subsection{What is the scaling parameter?}

Previously in this  Section  and in  Section \ref{second} we studied how  the effective perturbation at a given energy changes, when the cut-off  changes?
Alternatively, we can ask ourselves: How does the effective perturbation at a given  cut-off  change, when the energy changes?
That is we  are interested in
\begin{eqnarray}
\label{perturbat2}
dT_{\beta\alpha,ba}(\epsilon)\equiv T_{\beta\alpha,ba}(\epsilon+d\epsilon)-T_{\beta\alpha,ba}(\epsilon)
\end{eqnarray}
($d\epsilon$ can be of any sign). Looking at Eq. (\ref{perturbation2}) we understand that
\begin{eqnarray}
\label{perturba2}
&&dT_{\beta\alpha,ba}(\epsilon)=
\sum_{\gamma,c}V_{\beta\gamma,bc}V_{\gamma\alpha,ca}\sum_{-d\epsilon<\epsilon_{\bf q}<0}\frac{1}{\epsilon-\epsilon_{\bf q}}\nonumber\\
&&-\sum_{\gamma,c}V_{\gamma\alpha,bc}V_{\beta\gamma,ca}\sum_{0<\epsilon_{\bf q}<d\epsilon}\frac{1}{\epsilon_{\bf q}-\epsilon}\nonumber\\
&&=\rho\sum_{\gamma,c}\left[V_{\beta\gamma,bc}V_{\gamma\alpha,ca}
-V_{\gamma\alpha,bc}V_{\beta\gamma,ca}\right]\frac{d\epsilon}{\epsilon}.
\end{eqnarray}
(for the sake of definiteness we have chosen  $d\epsilon>0$).
Scaling equation we obtain by changing $V$ in the last line of Eq. (\ref{perturba2}) to $T$.
Thus we recover Eq. (\ref{scaling}),
 only this time
$\Lambda =|\epsilon|/D_0$.
Similarly, for the power law DOS, we recover Eq. (\ref{scaling8}), with
$\lambda =\left(|\epsilon|/D_0\right)^r$.

\section{The spin-anisotropic Kondo model}
\label{spin}

To see how the general scaling equation is applied let us consider the following spin-anisotropic  model  (summation  with respect to any repeated  Cartesian index is implied)
\begin{eqnarray}
\label{hamiltonian}
H=\sum_{{\bf k}\alpha}\epsilon_{\bf k}c_{{\bf k}\alpha}^{\dagger}c_{{\bf k}\alpha}
+\sum_{{\bf k}{\bf k}'\alpha\beta} J_{ij}S^i\sigma^j_{\alpha\beta}c_{{\bf k}'\alpha}^{\dagger}c_{{\bf k}\beta},
\end{eqnarray}
where
$S^x,S^y,S^z$ are the impurity  spin operators,
$\sigma^x,\sigma^y,\sigma^z$ are the Pauli matrices, and $J_{ij}$ is the anisotropic exchange coupling matrix.

The Hamiltonian (\ref{hamiltonian}) appears under many different names. Our opinion is that if we want to choose an eponymic name, the model should be called after T. Kasuya \cite{kasuya}. However, in line with the tradition we keep the name Kondo model, given because of  important contribution to the derivation and analysis of the model made by J. Kondo \cite{kondo0}.

Taking into account the commutation relations
\begin{eqnarray}
\label{spin2}
[S^i,S^j]=i\epsilon_{ijk}S^k,\;\;\;[\sigma^i,\sigma^j]=2i\epsilon_{ijk}\sigma^k,
\end{eqnarray}
where $\epsilon$ is Levi-Civita symbol, from Eq. (\ref{scaling9})   we obtain the scaling equation \cite{kogan}
\begin{eqnarray}
\label{scalinga}
\frac{d \widetilde{J}_{mn}}{d\lambda}=-\epsilon_{ikm}\epsilon_{jln}\widetilde{J}_{ij} \widetilde{J}_{kl}.
\end{eqnarray}
(In this Section we measure $J$ and $\tilde{J}$ in units of $r/2G$.)

We assume that the microscopic tensor $J_{ik}$ entering into the Hamiltonian (\ref{hamiltonian}) is  symmetric  (no spin-orbit interaction \cite{pletyukhov}). Analysis of the Hamiltonian in the presence of spin-orbit interaction see in Appendix \ref{orbit}. In this case the tensor
can  be reduced  to principal axes by rotation of the coordinate system, and it keeps it's
diagonal form  in the process of renormalization. So we can write down the interaction
in a diagonal in cartesian indices (though non explicitly rotation covariant) form
\begin{eqnarray}
\label{h}
V= J_{i}S^i\sigma^i.
\end{eqnarray}
The scaling equation for this interaction  is  \cite{kogan}
\begin{eqnarray}
\label{ggg}
\frac{d\widetilde{J}_i}{d\lambda}=-\widetilde{J}_j\widetilde{J}_k,
\end{eqnarray}
where $i,j,k$ are all different.

The general solution of Eq. (\ref{ggg}) (and, hence, of scaling equation) is  written in terms of elliptic functions  \cite{kogan}
\begin{eqnarray}
\label{amm}
J_{\alpha} &=&A\lambda\cdot\mathrm{ns}(A\lambda+\psi,k)\nonumber\\
J_{\beta}&=&A\lambda\cdot\mathrm{cs}(A\lambda+\psi,k)\\
J_{\gamma} &=&A\lambda\cdot\mathrm{ds}(A\lambda+\psi,k),\nonumber
\end{eqnarray}
where $\{\alpha,\beta,\gamma\}$ is an arbitrary permutation of  $\{x,y,z\}$. Using the language of geometry, we say that each flow line
 lies on the surface of a special cone
\begin{eqnarray}
\label{cone}
(1-k^2)J_{\alpha}^2+k^2J_{\beta}^2-J_{\gamma} ^2=0.
\end{eqnarray}
(For $k=0$ or $k=1$ the special cone is a pair of planes.)
Flow line passing through any given point with the coordinates $J_x^{(0)},J_y^{(0)},J_z^{(0)}$ is described by  Eq. (\ref{amm}) with $\alpha$ corresponding to the Cartesian component with maximal $|J^{(0)}|$, and
$\beta$ corresponding to the Cartesian component with minimal $|J^{(0)}|$.  The parameter $k$ for the cone, the flow line belongs to, is obviously found by substituting   $J_x^{(0)},J_y^{(0)},J_z^{(0)}$ into Eq. (\ref{cone}). Thus the whole phase space is divided into 3 domains, touching each other, each domain is defined by Eq. (\ref{cone})
with $k$ changing between $0$ and $1$ and corresponds to all the special cones with the axis along one of the Cartesian axes.

Detailed analysis of the solution (\ref{amm}) was presented earlier \cite{kogan}. Here we would like  to discuss only the finite asymptotics  of the solution  $\lambda\to 0$. For $\psi\neq 0,2K(k)$ (where $K$  is the complete elliptic integral of the
first kind), $(J_x,J_y,J_z)\to (0,0,0)$, which corresponds to a trivial fixed point. For
$\psi= 0$,  $(J_x,J_y,J_z)\to (1,1,1)$, and for
$\psi=2K(k)$,  $(J_{\alpha},J_{\beta},J_{\gamma})\to (-1,-1,1)$. The latter asymptotics correspond to four non-trivial fixed points
of the scaling equation
\begin{eqnarray}
\frac{1}{r}\frac{dJ_i}{d\ln\Lambda}&=&J_i-J_jJ_k.
\end{eqnarray}

\section{The  anisotropic CS model}
\label{gcs}

\subsection{The  CS model}

The CS model \cite{coqblin} is represented by the Hamiltonian
\begin{eqnarray}
\label{cs0}
&&H=\sum_m\epsilon c_m^{\dagger}c_m
+J\sum_{mm'} X_{mm'}c_{m'}^{\dagger}c_m        \nonumber\\
&&-\cancel{(J/N)\sum_{mm'}X_{mm}c_{m'}^{\dagger}c_{m'}},
\end{eqnarray}
where  quantum number $m$  changes from 1 to $N$. (To avoid cluttering, we  omit in the equations in this Section the wave vector indices.)
We changed sign of the constant $J$ with respect to the original paper \cite{coqblin}, so that the Hamiltonian (\ref{cs0}) would look like the Hamiltonian (\ref{hamiltonian}).
The last term in the R.H.S. of Eq. (\ref{cs0}) (and in similar equations further on) is crossed-out to show that it does not
give any contribution to the scaling equation, which for the Hamiltonian (\ref{cs0}) (and the constant DOS) has the form \cite{hewson}
\begin{eqnarray}
\frac{dJ}{d\ln\Lambda}=-N\rho J^2.
\end{eqnarray}
 For $N=2$  the model  coincides with the  spin-isotropic Kondo model.

\subsection{The  $XXZ$ CS model}

If we demand that interaction (\ref{h}) has $SU(2)$ symmetry, it takes the form
\begin{eqnarray}
\label{h3}
V= J\vec{S}\cdot\vec{\sigma}.
\end{eqnarray}
If we reduce the symmetry to $U(1)$,  interaction (\ref{h})   takes the form
\begin{eqnarray}
\label{h5}
V= J_x(S^x\sigma^x+S^y\sigma^y)+J_zS^z\sigma^z.
\end{eqnarray}
we will call such exchange interaction the $XXZ$ Kondo model.

Equation (\ref{h5}) can be written down using Hubbard operators
\begin{eqnarray}
\label{huhu}
V=J_x\left(X_{+-}c_{-}^{\dagger}c_{+}+X_{-+}c_{+}^{\dagger}c_{-}\right)\nonumber\\
+J_z\left(X_{++}c_{+}^{\dagger}c_{+}+X_{--}c_{-}^{\dagger}c_{-}\right)\nonumber\\
-\cancel{\frac{1}{2}J_z\left(X_{++}+X_{--}\right)\left(c_{+}^{\dagger}c_{+}+c_{-}^{\dagger}c_{-}\right)}.
\end{eqnarray}

Motivated by Eq. (\ref{huhu}) we suggest the following Hamiltonian for arbitrary $N$, which we for obvious reasons will call the  $XXZ$ CS model,
\begin{eqnarray}
\label{cs01}
&&H=\sum_m\epsilon c_m^{\dagger}c_m
+J_x\sum_{m\neq m'}X_{mm'} c_{m'}^{\dagger}c_m\nonumber\\
&&+J_z\sum_m X_{mm}c_{m}^{\dagger}c_m-\cancel{\frac{J_z}{N}\sum_{mm'}X_{mm}c_{m'}^{\dagger}c_{m'}}.
\end{eqnarray}
(An alternative motivation for introducing the model can be found in Appendix \ref{a3}.)
Further on in this Section we'll present the calculations only for the case of constant DOS,
and only the final results will be written down for the case of the power law DOS.

For the interaction (\ref{cs01}) scaling equation (\ref{scaling25}) becomes
\begin{eqnarray}
\label{scalingsc02}
\frac{dJ_x}{d\ln\Lambda} &=& -(N-2)\rho J_x^2-2\rho J_xJ_z\nonumber\\
\frac{dJ_z}{d\ln\Lambda} &=& -N\rho J_x^2.
\end{eqnarray}
For the case of isotropic CS model and for the case  $N=2$, Eq. (\ref{scalingsc02}) is  reduced to the  well established results \cite{hewson}.

\subsection{Integration of the scaling equation for the  $XXZ$ CS model}
\label{toy}

Dividing two equations in (\ref{scalingsc02}) by each other we obtain homogeneous differential equations of the first degree, which can be easily integrated
\begin{eqnarray}
\label{s22}
J_z &=& \frac{C}{\sqrt[N+2]{|w-1|^N\left|w+\frac{2}{N}\right|^2}}\\
\label{s23}
J_x &=& wJ_z,
\end{eqnarray}
where $C$ is an arbitrary constant.
Substituting the solution (\ref{s22}), (\ref{s23})   into Eq. (\ref{scalingsc02})  we obtain
\begin{eqnarray}
\label{scalingscgb}
\frac{dw}{d\ln\Lambda} = \frac{C N\rho w(w-1)\left(w+\frac{2}{N}\right)}{\sqrt[N+2]{|w-1|^N\left|w+\frac{2}{N}\right|^2}}.
\end{eqnarray}
(For $N=2$ Eq. (\ref{scalingscgb}) can be easily integrated in terms of elementary transcendental functions, otherwise in quadratures.)
Eq. (\ref{scalingscgb}) has three fixed points: $w=1$, $w=-2/N$, and $w=0$.
The first two are stable for $C>0$, the last one is  stable  for $C<0$.
(One should keep in mind that $\Lambda$ decreases in the process of evolution.)

At the phase plane with the coordinates $J_x,J_z$ the fixed point $w=1$  turns into the flow line $J_x=J_z$, and  the fixed point $w=-2/N$  turns into the flow line $J_x=-(2/N)J_z$. Both lines are attractors for $J_z>0$,
and  repellers (serving as the phase boundaries) for $J_z<0$. The $w=0$ fixed point turns into
the line of fixed points $J_x=0$, stable for $J_z<0$, and unstable for  $J_z>0$. (The half-line  $J_x=0$, $J_z>0$ serves as the phase boundary.) Notice that  $J_x=J_z=0$ is a degenerate fixed point \cite{arnold}.

Stable fixed point  correspond to  phases of Eq. (\ref{scalingsc02}), characterized by the asymptotic behaviour  of the flow lines. The part of the phase plane
$J_x>-(2/N) J_z,0$ is characterized by the attractor $J_x=J_z>0$, and will be called the Kondo phase I.
Notice that in this phase, the SU$(N)$ symmetry, which is absent for the microscopic Hamiltonian, is being recovered in the process of scaling.
The part $J_x<- J_z,0$ is characterized by the attractor
$J_x= -(2/N)J_z<0$, and will be called the Kondo phase II. In the part $-(2/N)J_z\geq J_x\geq J_z $ the flow lines are attracted to the fixed points at the axis $J_x=0$.

A flow line can reach the fixed point  $w=0$ only in the end of infinitely long evolution, which is a common situation for a fixed point. Hence Ising model is obtained only in the infrared limit.
However, in a Kondo phase a flow line reaches fixed point $w=1$ or $w=-2/N$ after finite evolution
(the R.H.S. of Eq. (\ref{scalingscgb}) being non-analytic function of $w$ at these points).
Singularity of $J_x,J_z$ at finite value of $\Lambda$ is another indication of the  limited applicability (in the Kondo phases) of the  truncated to  second order scaling equation.

To the general solution (\ref{s22}) we should add singular  solutions
\begin{eqnarray}
\label{hren5}
J_x&=&J_z=\frac{1}{N\rho \ln\Lambda+C_1}\nonumber\\
J_x&=&-\frac{2J_z}{N}=\frac{-1}{2\rho \ln\Lambda+C_2}.
\end{eqnarray}
Once again we see that  evolution starting  in one of the Kondo phases hits the singular point at finite value of $\ln\Lambda$.

Equation  (\ref{s22})     allows us to plot the flow diagram of the scaling equation, which  is presented on Fig. 1.  For the sake of definiteness we have chosen $N=4$.
Notice that the flow diagram is qualitatively similar to that of the $XXZ$ Kondo model \cite{hewson}.

\begin{figure}[h]
\vskip 1cm
\includegraphics[width= .9\columnwidth]{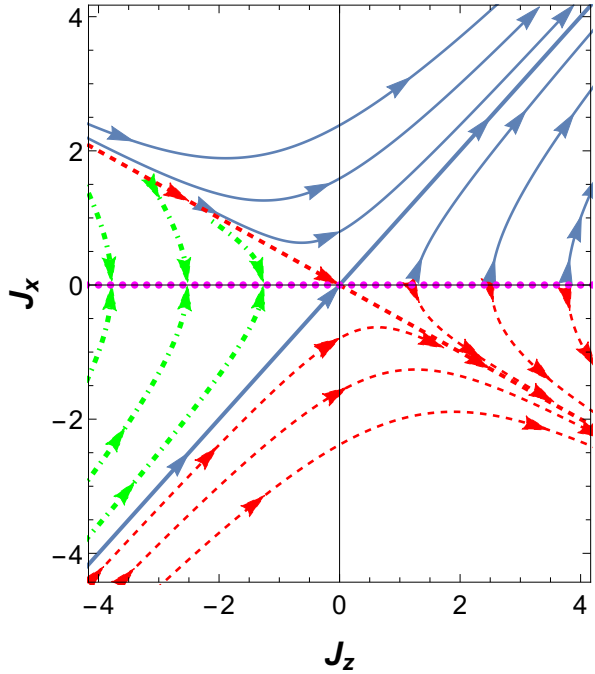}
\caption{(color online) Flow diagram for the scaling equation (\ref{scalingsc02}) for $N=4$. Solid (blue) flow lines belong to the Kondo phase I. Dashed (red) flow lines belong to the Kondo phase II.
Dashed dotted (green) flow lines belong to the Ising phase. Thick (magenta) dots on the $J_z$ axis are the  fixed points.}
 \label{XXZ}
\end{figure}

For the case of the power law DOS Eqs. (\ref{s22}) and (\ref{scalingscgb}) become
\begin{eqnarray}
\label{a11}
J_z = \frac{C\lambda}{\sqrt[N+2]{|w-1|^N\left|w+\frac{2}{N}\right|^2}}
\end{eqnarray}
and
\begin{eqnarray}
\label{a2}
r\frac{dw}{d\lambda} = \frac{CNG\omega(w-1)\left(w+\frac{2}{N}\right)}{\sqrt[N+2]{|w-1|^N\left|w+\frac{2}{N}\right|^2}}
\end{eqnarray}
(Eq. (\ref{hren5}) is changed similarly).

Eqs. (\ref{a11}) and (\ref{a2}) is a rigorous but a bit formal result. In particular, it demands some effort to
extract out of them the fixed points  of the original scaling equation
\begin{eqnarray}
\label{b11}
r\frac{dJ_x}{d\lambda} &=&rJ_x  -(N-2)G J_x^2-2G J_xJ_z\nonumber\\
r\frac{dJ_z}{d\lambda} &=&rJ_z -NG J_x^2,
\end{eqnarray}
which can be easily found by inspection of Eq. (\ref{b11}). The equation has a trivial fixed point $(J_x^*,J_z^*)=(0,0)$, which is stable for $r>0$ and unstable for $r<0$, and two semi-stable (critical) non-trivial fixed points
\begin{eqnarray}
J_x^* &=&\frac{2-N\pm\sqrt{(N-2)^2+8Nr^2}}{4NG}\nonumber\\
J_z^* &=&\frac{NG {J_x^*}^2}{r}.
\end{eqnarray}

\subsection{The  anisotropic CS model}
\label{XZ}

To formulate the general anisotropic CS model
let us return to the  spin-anisotropic Kondo model. The interaction can be written down using Hubbard operators
\begin{eqnarray}
\label{huhuhu}
V=\frac{J_x}{4}\left(X_{+-}+X_{-+}\right)\left(c_{+}^{\dagger}c_{-}+c_{-}^{\dagger}c_{+}\right)\nonumber\\
-\frac{J_y}{4}\left(X_{+-}-X_{-+}\right)\left(c_{+}^{\dagger}c_{-}-c_{-}^{\dagger}c_{+}\right)\nonumber\\
+J_z\left(X_{++}c_{+}^{\dagger}c_{+}+X_{--}c_{-}^{\dagger}c_{-}\right)\nonumber\\
-\cancel{\frac{J_z}{2}\left(X_{++}+X_{--}\right)\left(c_{+}^{\dagger}c_{+}+c_{-}^{\dagger}c_{-}\right)}.
\end{eqnarray}

Motivated by Eq. (\ref{huhuhu}) we suggest the following interaction for arbitrary $N$
\begin{eqnarray}
\label{cs01b}
V&=&\frac{J_x}{2}\sum_{m\neq m'}X_{mm'}\left(c_{m'}^{\dagger}c_{m}+c_{m}^{\dagger}c_{m'}\right)\nonumber\\
&+&\frac{J_y}{2}\sum_{m\neq m'}X_{mm'}\left(c_{m'}^{\dagger}c_{m}-c_{m}^{\dagger}c_{m'}\right)\nonumber\\
&+&J_z\sum_m X_{mm}c_{m}^{\dagger}c_m-\cancel{\frac{J_z}{N}\sum_{mm'}X_{mm}c_{m'}^{\dagger}c_{m'}}.
\end{eqnarray}
For this  interaction scaling equation (\ref{scaling25}) becomes
\begin{eqnarray}
\label{scalingsc01b}
\frac{dJ_x}{d\ln\Lambda}&=&-(N-2)\rho J_xJ_y-2\rho J_yJ_z\nonumber\\
\frac{dJ_y}{d\ln\Lambda}&=&-(N-2)\rho J_xJ_y-2\rho J_xJ_z\nonumber\\
\frac{dJ_z}{d\ln\Lambda}&=&-N\rho J_xJ_y.
\end{eqnarray}

Like in the previous Subsection we obtain after integration
\begin{eqnarray}
\label{s22b}
J_z&=&\frac{C_1}{\sqrt[N+2]{|w-1|^N\left|w+\frac{2}{N}\right|^2}}
=\frac{C_2}{\sqrt[N+2]{|v-1|^N\left|v+\frac{2}{N}\right|^2}}\nonumber\\\\
J_x &=& wJ_z,\hskip 2.5cm
J_y = vJ_z.
\end{eqnarray}
Analog of Eq. (\ref{scalingscgb}) can be  presented as
\begin{eqnarray}
\label{sb}
\frac{dw}{d\ln\Lambda} = \frac{C_1N\rho v w(w-1)\left(w+\frac{2}{N}\right)}{\sqrt[N+2]{|w-1|^N\left|w+\frac{2}{N}\right|^2}} .
\end{eqnarray}

Thus in the phase space with the coordinates $J_x,J_y,J_z$ there are 4 Kondo phases, each defined by the attractor of all the flow lines, given  by one of the vectors: $(1,1,1)$, $\left(-\frac{2}{N},-\frac{2}{N},1\right)$, $\left(-1,\frac{2}{N},-1\right)$, $\left(\frac{2}{N},-1,-1\right)$.
In addition there are 3 Ising phases, each defined by the fixed points of all the flow lines, lying  on one of the lines:
$J_x=J_y=0$, $J_x=J_z=0$, $J_y=J_z=0$.
Further analysis of the  phase diagram   we postpone until later.

For the case of the power law DOS Eqs. (\ref{s22b}) and (\ref{sb}) become
\begin{eqnarray}
J_z&=&\frac{C_1\lambda}{\sqrt[N+2]{|w-1|^N\left|w+\frac{2}{N}\right|^2}}
=\frac{C_2\lambda}{\sqrt[N+2]{|v-1|^N\left|v+\frac{2}{N}\right|^2}}\nonumber\\
\end{eqnarray}
and
\begin{eqnarray}
r\frac{dw}{d\lambda} = \frac{C_1NG v(w-1)\left(w+\frac{2}{N}\right)}{\sqrt[N+2]{|w-1|^N\left|w+\frac{2}{N}\right|^2}}.
\end{eqnarray}

In the end, notice that it would be interesting to apply the approach presented in this paper to the case, where  the exchange is influenced by Rashba \cite{sandler} or Dzyaloshinsky-Morya-Kondo interaction \cite{pletyukhov},  to the case when the DOS has a logarithmic singularity \cite{irkhin}, to the multi-channel Kondo model \cite{par}, and also to the problem of competition between the Kondo effect and RKKY interaction \cite{yudson}.
Another possibly interesting field for application of the presented approach  is the models when orbital and spin degrees of freedom of the impurity co-exist \cite{horvat2}.

\section{Conclusions}
\label{conclusions}

In the present contribution we  derive the poor man's scaling equation to the second order  for a general  model,
describing a quantum impurity with degenerate energy levels embedded into a gas of itinerant electrons, both for flat and for the power law DOS.
We  show how the obtained previously scaling equations for spin-anisotropic Kondo model follow from the general scaling equation.

We introduce  the  anisotropic  CS model, and  the $XXZ$ CS model as its particular case.
We   apply the general  scaling equation to derive scaling equations for  these  models.
We integrate analytically the scaling equation both in the  particular case of  the $XXZ$  CS model and in the general case of the anisotropic CS model.

\begin{acknowledgments}

The research leading to the results presented here was started during the author's  visit to
Max-Planck-Institut fur Physik komplexer Systeme, Dresden, continued during the author's visit to
DIPC, San Sebastian/Donostia, and  finalized during the author's visits to Keio University, Yokohama and National Cheng Kung University, Tainan.
The author  cordially thanks all the Institutions for the hospitality extended to him during
those and all his  previous visits.

The author is grateful to  N. Andrei, Y. Avishai, T. Costi, V. Golovach, K. Ingersent, V. Yu. Irkhin, T. Kimura, Min-Fa Lin, V. Meden, Y. Ohyama, M. Pletyukhov,  D. A. Ruiz-Tijerina, I. Tokatly, A. Weichselbaum,  O. M. Yevtushenko, G. Zarand,   and R. Zitko for valuable discussions.

\end{acknowledgments}

\begin{appendix}

\section{Real three-dimensional Lie algebras}
\label{three}

Every Lie Algebra over a real three-dimensional vector space is isomorphic with
one of the following algebras appearing in the Table \ref{table} \cite{patterson}:
\begin{table}
\begin{tabular}{|l|l|l|l|l|}
\hline
$L_1$: & $f^1_{23}=1$ & $f^2_{31}=1$ & $f^3_{12}=1$ & \\
$L_2$: & $f^1_{23}=1$ & $f^2_{31}=1$ & $f^3_{12}=-1$ & \\
$L_3(\alpha)$: & $f^1_{23}=1$ & $f^2_{23}=\alpha$ & $f^2_{31}=1$ & $(\alpha\geq 0)$ \\
$L_4(\alpha)$: &$f^1_{23}=1$ & $f^2_{23}=\alpha$ & $f^2_{31}=-1$ & $(\alpha\geq 0)$ \\
$L_5$: & $f^2_{23}=1$ &$f^1_{31}=-1$ &  & \\
$L_6$: & $f^1_{23}=1$ &  &  & \\
$L_7$: & $f^2_{23}=1$ &  &  & \\
\hline
\end{tabular}
\caption{Possible real three-dimensional Lie algebras. In each case the remaining structure constants $f^p_{st}$ except those appearing in the table are zero; we have omitted zero algebra.}
\label{table}
\end{table}
The algebra $L_1$ is isomorphic to $su(2)$ algebra; the algebra $L_2$ gives the same scaling equation as $L_1$.
We will write down explicitly only scaling equation (\ref{scaling26}) for the algebras $L_3,L_4$.
Such equation (we assume $\rho=1$) is
\begin{eqnarray}
\label{alpha}
\frac{dc_1}{d\ln\Lambda}&=&-c_2c_3 \nonumber\\
\frac{dc_2}{d\ln\Lambda}&=&-c_1c_3-\alpha^2c_2c_3 \nonumber\\
\frac{dc_3}{d\ln\Lambda}&=&0.
\end{eqnarray}
Equation (\ref{alpha}) can be easily solved in terms of exponential functions.
Scaling equations for
the algebras $L_5,L_6,L_7$ are even simpler.

It is worth pointing to the Lie groups $g_3(\alpha)$ and $g_4(\alpha)$, which have $L_3$ and $L_4$ as their Lie algebras.(The group $g_1$ can be represented by the group of rotations
which keep the form $f_1=x_1x_2+y_1y_2+z_1z_2$ invariant,  and the group $g_2$ - by the group of "rotations"
which keep the form $f_2=x_1x_2+y_1y_2-z_1z_2$ invariant). It is shown in Ref. \cite{patterson} that the  group $h_3(\alpha)$ can be represented by the matrices of the form
\begin{eqnarray}
\left(\begin{array}{ccc} f'(z) & f(z) & 0\\
                         -f(z) & f'(z)-\alpha f(z) & 0 \\
                         y & x & 1
                         \end{array}\right),
\end{eqnarray}
where $f(z)=\beta^{-1}\exp\left(\frac{1}{2}\alpha z\right)\sinh(\beta z)$ $(\beta=\frac{1}{2}\sqrt{\alpha^2-4})$ if $\alpha>2$, $f(z)=z\exp x$
if $\alpha=2$ and  $f(z)=\gamma^{-1}\exp\left(\frac{1}{2}\alpha z\right)\sin(\gamma z)$ $(\gamma=\frac{1}{2}\sqrt{4-\alpha^2})$ if $0\leq\alpha<2$.
The  group $h_4(\alpha)$ can be represented by the matrices of the form
\begin{eqnarray}
\left(\begin{array}{ccc} f'(z)-\alpha f(z) & f(z) & 0\\
                         f(z) & f'(z) & 0 \\
                         x & y & 1
                         \end{array}\right),
\end{eqnarray}
where $f(z)=\delta^{-1}\exp\left(\frac{1}{2}\alpha z\right)\sinh(\delta z)$ $(\delta=\frac{1}{2}\sqrt{\alpha^2+4})$.

\section{The Dzyaloshinskii-Moriya-Kondo interaction}
\label{orbit}

In the presence of spin-orbit interaction the Hamiltonian (\ref{hamiltonian}) can be written as \cite{pletyukhov}
\begin{eqnarray}
\label{hamil}
H=\sum_{{\bf k}\alpha}\epsilon_{\bf k}c_{{\bf k}\alpha}^{\dagger}c_{{\bf k}\alpha}+\sum_{{\bf k}{\bf k}'\alpha\beta} J_{ij}S^i\sigma^j_{\alpha\beta}c_{{\bf k}'\alpha}^{\dagger}c_{{\bf k}\beta}\nonumber\\
+\sum_{{\bf k}{\bf k}'\alpha\beta} \vec{D}
\cdot[\vec{S}\times\vec{\sigma}_{\alpha\beta}]c_{{\bf k}'\alpha}^{\dagger}c_{{\bf k}\beta},
\end{eqnarray}
where we keep the matrix $J_{ij}$ symmetric but introduce the  Dzyaloshinskii-Moriya \cite{dzya,moriya} (DM) interaction; $\vec{D}$ is the DM vector.
Mathematically, while writing down  Eq. (\ref{hamil}) we just presented arbitrary asymmetric matrix as a sum of symmetric and antisymmetric ones.
We just mention here, that if we impose $U(1)$ symmetry, in an appropriate coordinate system the matric $J_{ij}$ will be written as $J_{ij}$=diag$(J_x,J_x,J_z)$, and the vector $\vec{D}$ as  $\vec{D}=(0,0,D_z)$  \cite{pletyukhov}.

\section{The CS model revisited}
\label{a3}

To warm up, let us start from the Kondo model.
Historically, first isotropic case was studied, then the $XXZ$ model, and then the completely  anisotropic model.
Let us mentally inverse the process, and try to understand how the  $XXZ$ model could have appeared as
a renormalizable particular case of the  Hamiltonian (\ref{h}).

First, a general mathematical statement. Because the scaling equation (\ref{scaling26}) obviously keeps the symmetry of the Hamiltonian,
we can obtain renormalizable particular case of the general Hamiltonian (\ref{product}) by imposing on it some symmetry, and
considering the most general Hamiltonian compatible with the chosen symmetry.

Now back to Kondo model. Let us impose on  the Hamiltonian (\ref{h})
the U$(1)$ symmetry.
The group U$(1)$ has only 3 bilinear invariants, which all appear in Eq. (\ref{huhu}). (We took into account additionally symmetry of any spin Hamiltonian with respect to inversion, which in  our case means symmetry with respect to interchange of $+$ and $-$.) The condition $<V>=0$
demands the relation between the coefficients of two of them,  and
we recover  Eq. (\ref{huhu}).

Now comes the  CS model. It can be obtain from the Hamiltonian
\begin{eqnarray}
\label{cs}
&&H=\sum_m\epsilon c_m^{\dagger}c_m
+\sum_{mm'} J_{m'm}X_{mm'}c_{m'}^{\dagger}c_m\nonumber\\
&&-\sum_{mm'}J'_{m'm}X_{m'm'}c_m^{\dagger}c_m,
\end{eqnarray}
by imposing upon upon it SU$(N)$ symmetry.   In fact, if we introduce matrices  $\hat{C}$ and $\hat{X}$, with the matrix elements $c_{m'}^{\dagger}c_m$ and  $X_{m'm}$ respectively,
the interaction should contain only the bilinear combinations  of the  matrix elements  that are invariants of the symmetry group.
For the group SU$(N)$ there are only two of them:  Tr$( \hat{C}\cdot\hat{X})$ and  Tr$\hat{C}\cdot$Tr$\hat{X}$. If we impose additional condition \cite{coqblin}
$<V>=0$,
to remove from the Hamiltonian the direct (potential) term, we exactly reproduce Eq. (\ref{cs0}).

It is tempting to assume that the  Hamiltonian (\ref{cs01}) can be obtained by imposing on the Hamiltonian (\ref{cs})
the symmetry U$(N-1)$  (the maximal subgroup of SU$(N)$  \cite{antoneli}). However we are unable to either  prove  or  disprove this assumption.

In general,  the group theory approach suggests the way to get the particular cases of the Hamiltonian (\ref{cs}), which are renormalizable. We should just choose some subgroup of SU$(N)$ as the symmetry group of the Hamiltonian (\ref{cs}), and write down a general linear combination of all the bilinear invariants.

\end{appendix}

\end{document}